\documentclass[usletter]{article}
\pdfoutput=1            
\usepackage{color}
\usepackage{graphicx}
\usepackage{xspace}
\usepackage{authblk}

\def\xrootd{XRootD\xspace}

\usepackage[pdftex,bookmarks]{hyperref}
\hypersetup{
    colorlinks=false,
    linktoc=section,
    linkcolor=black,
    citecolor=blue,
    urlcolor=blue,
    filecolor=black,
    pdfpagemode=UseOutlines,
}

\begin{document}
\title{HEP-FCE Working Group on Libraries and Tools}

\author{Anders Borgland}
\affil{SLAC}

\author{Peter Elmer}
\affil{Princeton}

\author{Michael Kirby}
\affil{FNAL}

\author{Simon Patton}
\affil{LBNL}

\author{Maxim Potekhin}
\affil{BNL}

\author{Brett Viren}
\affil{BNL}

\author{Brian Yanny}
\affil{FNAL}

\date{December 19, 2014}
\maketitle

\tableofcontents
\pagebreak
\section{Introduction}

The High-Energy Physics Forum for Computational Excellence (HEP-FCE)
was formed by the Department of Energy as a follow-up to a recent
report from the Topical Panel on Computing\cite{topicalpanel} and the
associated P5 recommendation\cite{p5}.
It is a  pilot project distributed across the DOE Labs.  
During this initial incubation period the Forum is to develop a plan
for a robust, long-term organization structure and a functioning web
presence for forum activities and outreach, and a study of hardware
and software needs across the HEP program.

Three FCE Working Groups have been constituted as a limited-duration
effort to list and prioritize computational activities across the
three HEP frontiers.  This report summarizes the deliberations of the
\textbf{HEP-FCE Software Libraries and Tools Working Group}. The
charge to the working group includes topics such as:

\begin{itemize}
\item Code management utilities
\item Build/release/scripting/testing tools
\item Documentation tools
\item Graphics packages
\item General purpose libraries (I/O, statistical analysis, linear algebra)
\item  Data management and transfer tools
\item  Workflow and Workload management
\end{itemize}

The other two working groups focus on Systems (computing, data and networking)
and Applications (the software itself and its distribution).   
Even after narrowing to just software libraries and tools, the breadth and
depth of work relevant to HEP is far too extensive to cover in this
document.  Instead of attempting to be comprehensive the working group
has considered only a sampling, hopefully representative, of the
possible projects and areas.  Omissions are not intended to be
interpreted as positive or negative reflections on those projects or
areas.

In the following sections we give this working group's ``vision'' for
aspects and qualities we wish to see in a future HEP-FCE.  We then
give a prioritized list of technical activities with suggested scoping
and deliverables that can be expected to provide cross-experiment
benefits.  The remaining bulk of the report gives a technical survey of some
specific ``areas of opportunity'' for cross-experiment benefit in the
realm of software libs/tools.  This survey serves as support for the
vision and prioritized list.  For each area we describe the ways that
cross-experiment benefit is achieved today, as well as describe known
failings or pitfalls where such benefit has failed to be achieved and
which should be avoided in the future.  For both cases, we try to give
concrete examples.  Each area then ends with an examination of what
opportunities exist for improvements in that particular area.

\section{Vision Statement}

The working group members choose to take this opportunity to describe
what we would like to see materialize to improve beneficial
cross-experiment software development and use.

\subsection{HEP Software Foundation}

The HEP Software Foundation\cite{hsfweb} (HSF) is forming at the time
of the writing of this report.  It shares many of the goals of the
working group. The eventual FCE should work with HSF in mutually
beneficial ways.

\subsection{Community}

One or more nuclei, around which the HEP software community can
consolidate to facilitate better cross-experiment software development
and use are needed.  For a community nucleus to be beneficial it must be created
with understanding and sensitivity to how the members of the community
naturally work.  Any Internet-based forum that lacks this will simply
be left unused.  Specific aspects of a community nucleus may include:

\begin{itemize}
\item Email lists specific to the various areas of software.  They should be
  newly created and hosted if missing or centrally advertised if already existing.
\item A ``market'' where experiments can go to advertise their needs,
  solicit ideas from a broad swath of the community and form 
  collaborative efforts to satisfy these needs.
\item An ``incubator'' where novel development or improvement and
  generalization of existing software can be discussed and directed
  through community involvement and when possible contributed to by
  available community members.  A component of this incubator would
  include developing funding proposals.  (See also section~\ref{sec:support}.)
\item A ``knowledge-base'' filled with collaboratively produced content
  that includes items such as:
  \begin{itemize}
  \item summary information on individual software libraries and tools
    which are considered useful by and for the community.
  \item information about experiments, projects and other
    organizations and associations with the software they use.
  \item contact information on community individuals and the software
    and experiments in which they are involved.
  \item an indexed archive of software documentation, publications,
    ``how to'' guides maintained directly in the knowledge-base or 
    as external links to existing ones.
  \end{itemize}
\end{itemize}

\noindent These online resources should be open for viewing by the world and
indexed by Internet search engines to facilitate information
discovery.

The community should receive periodic reports from the leaders of
these organizational efforts as part of existing HEP software and
computing conferences.

\subsection{Support}
\label{sec:support}

We would like to see a method where novel effort beneficial to
multiple experiments can be funded.  The desired nature of such
funding is described here.

Funding proposals may come out of the ``incubator'' described above but need not.
The effort may be toward making an existing software more generally
beneficial to multiple experiments
or toward providing new development or an improved (re)implementation that is 
considered beneficial for and by the community.  This may include factoring
general parts from experiment-specific libs/tools, adoption of one
experiment's software by another or a ground up novel design and
implementation which provides a more general instance of existing
experiment-specific software.

For such proposed effort, a balance must be struck between addressing
the needs of one (or more) specific experiment and having the
development of a proper, general solution not be rushed by those
immediate needs.  Proposals may be born from the needs of a specific
experiment but should be judged by how useful the proposed solution
will be to experiments in the future or the larger existing community.
The effort should not be negatively judged solely due to not being
concluded in time to be used by the originating experiment.

Such proposals should be specific to the development of some software
deliverable.  They should include expected mechanisms to provide long
term support and improvement but funding for any ongoing effort should
not be part of such proposals.  Proposals may be for effort to
accomplish some specific novel deliverable or one which augments or
improves existing software.

Software resulting from such supported effort must be made available
to the HEP community on terms consistent with an  established
Free Software or Open Source license such as the GNU GPL or BSD licenses.

\pagebreak
\section{Prioritized Recommendations}

\subsection{Cross-Experiment Effort}

This section presents priority recommendations that the working group
gives to the HEP-FCE with an understanding that they are of interest
to funding entities.

\begin{enumerate}
\item The goals of the HEP Software Foundation are
  largely overlapping with HEP-FCE.  The two groups can be mutually
  beneficial.  HSF is forming a general, grassroots organization while
  HEP-FCE potentially can fund specifically targeted efforts.  HSF can
  be a source of requirements, working groups, review panels, project
  proposals and expertise all of which HEP-FCE can beneficially
  leverage to carry out its goals.  It is recommended that HEP-FCE
  continue to participate in HSF formation activities and look into
  how the two groups might partner on specific actions.

\item Various detailed ``opportunities'' listed in the following
  survey sections call out the need for further work to be carried out
  in some detail by technical working groups. These are needed to
  better understand the nature of a specific problem shared across
  many experiments, formulate requirements for and in some cases
  design and implement solutions.  The HEP-FCE should organize such
  working groups from suitable expertise from the HEP software community.

\item Packages or frameworks (or significant subsets) which have proven 
popular (used by more than one experiment) and useful should be considered for
cross-experiment support, especially in terms of providing support for
easy adoptability (setup and install by other experiments, on other 
O/S platforms) and documentation (detailed guides and non-experiment-specific manuals).

\end{enumerate}

\subsection{Effort by Experiments}

Throughout the following survey sections, a number of best practices
and pitfalls relevant to the development and use of software libraries
and tools by individual experiments were identified.  First, some
generalities were identified:

\begin{enumerate}

\item New experiments should not underestimate the importance of
  software to their success.  It should be treated as a major
  subsystem at least on par with other important aspects such as
  detector design, DAQ/electronics, civil construction, etc.

\item Experiments should understand the pitfalls listed in
  section~\ref{subsec:pitfalls}).  New experiments should plan and
  implement mechanisms to avoid them and existing experiments should
  reflect on which ones may apply and develop ways to address them.
  Likewise, the best practices listed in
  section~\ref{subsec:bestpractices}) should be considered.  New
  experiments should attempt to follow them and if practical and
  beneficial, existing experiments should seek to make the changes
  needed to implement them.

\item New and existing experiments should join the effort to organize
  and otherwise supply representative members to the HEP Software
  Foundation.

\end{enumerate}

\noindent The remaining sections below contain surveys of select areas
of software libraries and tools in HEP.  For each we list a summary of
aspects that make for success in the associated area.

\begin{itemize}

\item Aspects of successful Event Processing Software Frameworks include: 
those with flexible (possibly hierarchical) notions of ``events'', 
those that are easily adopable by new experiments, are well-documented, 
have dynamically configuable (possibly scriptable) configuration 
parameter sets and are modular and efficient (allow C++ like modules 
for low level 
operations combined with a scripting layer like python for flexible 
higher level control).

\item Aspects of successful Software Development tools include:
those that follow licence-free availablity and free-software distribution models, 
those that include code repositories, build systems that work on a variety of platforms 
with a small number of clearly defined base element dependencies 
(i.e. C compiler, compression library, specific version of python) and those with release 
configuration systems with versioning that understand a variety of 
platforms; those that support automatic continuous integration and 
regression testing; those that have open documentation updating and  
bug-reporting and tracking.

\item Aspects of successful Data Management tools include:
those that are inherently modular and avoid tight couplings to either 
specific technologies or to other parts of the computing ecosystem, in 
particular to the Workload Management System and the Metadata Catalogs;  
those that, while being complex and originallly developed for large scale operations 
at for example the LHC, may be simplified and downscaled for use by smaller experiments 
with minimal manpower and technical expertise. 

\item Aspects of successful Workflow and Workload Management tools include:
those that understand the distinction between and support flexible, efficient 
interaction between workflow, workload, and data management aspects 
of a system; those that make efficient use of resources (CPU, RAM-memory, 
disk, network, tape) for processing in parallel;  those that allow 
granular, multi-level monitoring of status; those that handle error 
cases effectively and inclusively; those that are properly scaled 
to the size of the experiment.

\item Aspects of successful Geometry Information Management tools include:
those that follow or set widely used standards for representation of
geometric information; those that follow standards for visualization.

\item Aspects of successful Conditions Database tools include:
those that allow standardized, experiment-wide access to representative 
or specific event conditions so that realistic simulations or statistics
can be generated by users without detailed knowledge of detectors or specific
event.

\end{itemize}

\pagebreak
\section{Survey of Current Landscape}


This section presents a general overview of the current landscape of
HEP libraries and tools.  
First we list general patterns that run counter to cross-experiment sharing.
%
%
Secondly, we give a prioritized list of activities and where involvement 
by the HEP-FCE could be beneficial.

\subsection{Forces Counter to Cross-experiment Software}
\label{subsec:pitfalls}

Sharing software libraries and tools between experiment more
frequently than is currently done is expected, by the group, to
increase overall productivity.  Independent of cross-experiment
sharing, designing and implementing software in a more general manner
is expected to be beneficial.  The working group
identified some reasons why such general use software is not as
predominant as it could be.

\subsubsection{Up-front Effort}

Designing and implementing software to solve a general problem instead
of the specific instance faced by one experiment can take more effort
initially.  Solving ``just'' the problem one immediately faces is
cheaper in the immediate time scale.  If the problem is short lived
and the software abandoned this strategy can be a net-benefit.  What
is more often the case, fixes to new problems compound the problem and
the software becomes either brittle and narrowly focused, increasingly
difficult to maintain, and ever less able to be extended.

\subsubsection{Lack of Expertise}

Physicist have always been multidisciplinary, covering all aspects of
an experiment from hardware design, bolt turning, operations, project
management, data analysis and software development.  As data rates
increased, algorithms become more complex and networking, storage and
computation technology more advanced, the requirements for a Physicist to
be a software developer have become more challenging to meet while
maintaining needed capabilities in the other
disciplines.  As a consequence, some experiments, especially the smaller
ones, lack the software expertise and knowledge of what is available
needed to develop general software solutions or adopt existing ones.
This leads to the same result of solving ``just'' the immediate
problem and associated consequences described above.

\subsubsection{Ignoring Software Design Patterns}

A specific example of lack of expertise manifests in developers who
ignore basic, tried and true software design patterns.  This can be
seen in software that lacks any notion of interfaces or layering
between different functionality.  Often new features are developed by
finding a spot that ``looks good'' and pasting in some more code to
achieve an immediate goal with no understanding of the long-term consequences.
Like the ``up-front'' costs problem, this strategy is often rewarded as the
individual produces desired results quickly and the problem that this change
causes is not made apparent until later.

\subsubsection{Limited Support}

Some experiments have a high degree of software expertise.  These
efforts may even naturally produce software that can have some
cross-experiment benefit.  However, they lack the necessary ability to
support their developers to make the final push needed to offer that
software more broadly.  In many cases they also do not have the ability
to assure continued support of the software for its use by others.
In the best cases, some are able to provide support on a limited or best effort basis.  
While this helps others adopt the software it still leaves room
for improvements.  A modest amount of expert time can save a large amount of time of many novices.

\subsubsection{Transitory Members}

Many software developers in an experiment are transitory.  After
graduate students and post-docs make a contribution to the software
development and the experiment in general they typically move on to
other experiments in the advancement of their careers.  In part, this migration
can help disseminate software between experiments but it also
poses the problem of retaining a nucleus of long-term knowledge and
support around the software they developed.

\subsubsection{Parochial View}

In some cases, beneficial software sharing is hampered by experiments,
groups, labs, etc which suffer from the infamous ``not invented here''
syndrome.  A parochial view leads to preferring solutions to come from
within the unit rather than venturing out and surveying a broader
landscape where better, more general solutions are likely to be found.
Parochialism compounds itself by making it ever more difficult for
motivated people to improve the entrenched systems by bringing in more 
general solutions.

\subsubsection{Discounting the Problem}

There is a tendency with some Physicists to discount software and
computing solutions.  The origin of this viewpoint may be due to the
individual having experience from a time where software and computing
solutions were indeed not as important as they are now.  It may also
come as a consequence of that person enjoying the fruits of high
quality software and computing environments and being ignorant of the
effort needed to provide and maintain them.  Whatever the origin,
underestimating the importance of developing quality software tools leads
to inefficiency and lack of progress. 

\subsubsection{Turf Wars}

Software development is a personal and social endeavor.  It is natural
for someone who takes pride in that work to become personally attached
to the software they develop.  In some cases this can cloud judgment
and lead to retaining software in its current state while it may be
more beneficial to refactor or discard and reimplement.  What are
really prototypes can become too loved to be replaced.

\subsubsection{Perceived Audience and Development Context}

The group made the observation that cues from the audience for the
software and the context in which it is developed lead to shifts in
thinking about a software design.  For example, resulting designs tend to be more narrowly
applicable when one knows that the code will be committed to a private
repository only accessible by a single collaboration.  On the other hand, 
when one is pushing commits to a repository that is fully accessible
by a wide public audience one naturally thinks about broader use cases and 
solutions to more general problems.

\subsubsection{Disparate Communications}

Different experiments and experiment-independent software projects
have differing means of communicating.  Technical support, knowledge
bases, software repositories, bug trackers, release announcements are
all areas that have no standard implementation.  Some groups even
have multiple types of any of these means of communication.
Independent of this, different policies mean that not all information
may be publicly available.  These all pose hurdles for the sharing of 
software between groups.

\subsubsection{Design vs. Promotion}

For general purpose software to be beneficial across multiple
experiments it needs at least two things.  It needs to be well
designed and implemented in a way that is general purpose.  It
also needs to be promoted in a way so that potential adopters learn of
its suitability.  Often the set of individuals that excel at the
former and excel at the latter have little overlap.

\subsubsection{Decision Making}

An experiment's software is no better than the best software expert
involved in the decision making process used to provide it. And it's often worse.
Decision making is a human action and as such it can suffer from being
driven by the loudest argument and not necessarily the one most sound.
Many times, choices are made in a vacuum lacking suitable opposition.
At times they are made without a decision-making policy and procedures in place or
ignored if one exists, or if followed, without sufficient information
to make an informed decision.  Politics and familiarity can trump
rationality and quality.

\subsubsection{Getting off on the wrong foot}

There is often no initial review of what software is available when a new experiment begins.
Frequently a Physicist charged with software duties on an experiment will 
jump in and begin to do things the way that they were done in their 
last project, thus propagating and baking in inefficencies for another generation.  
No time will be spent to see what has changed since an earlier experiment's 
software design, and whole evolutions in ways of thinking or recently available
tools updates may be missed.

\subsection{Best Practices for Experiments}
\label{subsec:bestpractices}

\subsubsection{Look around}

New experiments should survey and understand the current state of the
art for software libraries and tools (and Systems and Applications as
covered by the other two working groups).  Periodically, established
experiments should do likewise to understand what improvements they
may adopt from or contribute to the community.  Experts from other
experiments should be brought in for in-depth consultation even in
(especially in) cases where the collaboration feels there is
sufficient in-house expertise.

\subsubsection{Early Development}

There are certain decisions that if made early and implemented can
save a lot of effort in the future.  Experiments should take these
seriously and include them in the conceptual and technical design
reports that are typically required by funding agencies.  These
include the following:

\begin{description}
\item[data model] Detailed design for data model schema covering the
  stages of data production and processing including: the output of
  detector simulation (including ``truth'' quantities), the output of
  ``raw'' data from detector DAQ and the data produced by and used as
  intermediaries in reconstruction codes.

\item[testing] Unit and integration testing methods, patterns and
  granularity.  These should not depend on or otherwise tied to other
  large scale design decisions such as potential event processing
  software frameworks.  

\item[namespaces] Design broad-enough namespace rules (unique filenames, 
event numbering conventions, including re-processed event version tags) to 
encompass the entire development, operations and legacy aspects of the 
experiment, which may span decades in time and have worldwide distributed 
data stores.  Filenames, or, in larger experiments, the meta-system 
which supports file access and movement, should have unique identifiers 
not just for given events or runs at a single location, but even if 
a file is moved and mixed with similar files remotely located 
(i.e. filename provenance should not rely upon directory path for 
uniqueness).  One should be able to distinguish development versions 
of files from production versions.  
If the same dataset is processed multiple times, the filenames or 
other metadata or provenance indicators should be available which 
uniquely track the processing version.  The same goes for code:  
software versions must be tracked clearly and comprehensively across the 
distributed experiment environment (i.e. across multiple 
institutions, experiment phases and local instances of repositories).

\item[scale] Understand the scale of complexity of the software, its
  development/developers.  Determine if an event processing framework
  is needed or if a toolkit library approach is sufficient or maybe if
  ad-hoc development strategies are enough.

\item[metadata] Determine what file metadata will be needed across the
  entire efforts of the collaboration.  Include raw data and the
  requirements for its production as well as simulation and processed
  data.  Consider what file metadata will be needed to support large
  scale production simulation and processing.  Consider what may be
  needed to support ad-hoc file productions by individuals or small
  groups in collaboration.

\end{description}

\subsection{Areas of Opportunity}

Each of the following sections focus on one particular \textit{area of opportunity} to make improvements in how the community shares libs/tools between experiments.  In each area of opportunity we present:

\begin{itemize}
\item A description of the area.
\item A number of case studies of existing or past software libraries and tools including concrete examples of what works and what does not.
\item Specific aspects that need improvement and an estimation of what efforts would be needed to obtain that.
\end{itemize}

\pagebreak
\section{Event Processing Software Frameworks}

\subsection{Description}

A software framework abstracts common functionality expected in some
domain.  It provides some generic implementation of a full system in
an abstract way that lets application-specific functionality to be
added through a modular implementation of framework interfaces.

Toolkit libraries provide functionality addressing some domain in a
form that requires the user-programmer to develop their own
applications.  In contrast, frameworks provide the overall flow
control and main function requiring the user-programmer to add
application specific code in the form of modules.

In the context of HEP software, the terms ``event'' and ``module" are often overloaded and poorly defined.
In the context of software frameworks, an ``event'' is a unit of data whose scope is dependent
on the ``module'' of code which is processing.  In the context of a
code module that generates initial kinematics, an event is the
information about the interaction.  In a module that simulates the
passage of particles through a detector, an event may contain all
energy depositions in active volumes.  In a detector electronics
simulation, it may contain all signals collected from these active
volumes.  In a trigger simulation module, it would be all readouts of
these signals above some threshold or other criteria.  At this point,
data from real detectors gain symmetry with simulation.  Going further,
data reduction, calibration, reconstruction and other analysis modules
each have a unique concept of the ``event'' they operate on.
Depending on the nature of the physics, the detector, and the follow-on
analysis, every module may not preserve the multiplicity of data.  For
example, a single interaction may produce multiple triggers, or none.

With that description, an event processing software framework is
largely responsible for marshalling data through a series (in general a
directed and possibly cyclic graph) of such code modules which then mutate the data.  To support
these modules the framework provides access to external services such
as data access, handle file I/O, access
to descriptions of the detectors, provide for visualization or
statistical summaries, and databases of conditions for applying
calibrations. 
The implementation of these services may be left up to the experiment
or some may be generically applicable.
How a framework marshals and associates data together as an event is
largely varied across different HEP experiments and may be unique for
a given data collection methodology (beam gate, online trigger, raw
timing, etc).


\subsection{Gaudi}

The Gaudi event processing framework\cite{gaudi_lhcb,gaudi_atlas} provides a comprehensive set of
features and is extensible enough that it is suitable for a wide
variety of experiments.  It was conceived by LHCb and adopted by ATLAS
and these two experiments still drive its development.  It has been
adopted by a diverse set of experiments including HARP, Fermi/GLAST,
MINER$\nu$A, Daya Bay and others.  The experience of Daya Bay is
illuminating for both Gaudi specifically and of more general issues of
this report.

First, the adoption of Gaudi by the Daya Bay collaboration 
was greatly helped by the support from the LHCb
and ATLAS Gaudi developers.  Although not strictly their
responsibility, they found the time to offer help and support to this and the other
adopting experiments.  Without this, the success of the adoption would have been uncertain
and at best would have taken much more effort.  Daya Bay recognized
the need and importance of such support and, partly selfishly, formed
a mailing list\cite{gauditalk} and solicited the involvement of Gaudi developers from many
of the experiments involved in its development and use.  It became a forum that more efficiently spread
beneficial information from the main developers.  It also offloaded
some support effort to the newly minted experts from the other experiments so that they could help
themselves.

There were, however areas that would improve the adoption of Gaudi.
While described specifically in terms of Gaudi they are general in nature.
The primary one would be direct guides on how to actually adopt it.
This is something that must come from the community and likely in
conjunction with some future adoption.  Documentation on Gaudi itself
was also a problem particularly for Daya Bay developers where many of the basic
underlying framework concepts were new.  Older Gaudi design documents and
some experiment-specific ones were available but they were not always accurate
nor focused on just what was needed for adoption.  Over time, Daya Bay produced it's own
Daya Bay-specific documentation which unfortunately perpetuates this
problem.

Other aspects were beneficial to adoption.  The Gaudi build system,
based on CMT\cite{cmt} is cross platform, open and easy to port.  It has layers
of functionality (package build system, release build system, support
for experiment packages and ``external'' ones) but it does not require a full
all-or-nothing adoption.  It supports a staged adoption approach that
allowed Daya Bay to get started using the framework more quickly.

The importance of having all Gaudi source code open and available can
not be diminished.  Also important was that the Gaudi developers included
the growing community in the release process.

While Gaudi's CMT-based package and release build system ultimately
proved very useful, it hampered initial adoption as it was not commonly
used widely outside  of Gaudi and the level of understanding required was high.  
It is understood that there is now a
movement to provide a CMake based build system.  This may alleviate
this particular hurdle for future adopters as CMake is widely used both inside and outside HEP projects.

Finally, although Gaudi is full-featured and flexible it did not come
with all needed framework-level functionality and, in its core, does
not provide some generally useful modules that do exist in experiment code repositories.  In particular, Daya Bay
adopted three Gaudi extensions from LHCb's code base.  These are
actually very general purpose but due to historical reasons were not
provided separately.  These were GaudiObjDesc (data model definition),
GiGa (Geant4 interface) and DetDesc (detector description).  Some
extensions developed by other experiments were rejected and in-house
implementations were developed.  In particular, the extension that
provided for file I/O was considered too much effort to adopt.  The
in-house implementation was simple, adequate but its performance was
somewhat lacking.

One aspect of the default Gaudi implementation that had to be modified
for use by Daya Bay was the event processing model.  Unlike collider
experiments, Daya Bay necessarily had to deal with a non-sequential,
non-linear event stream.  Multiple detectors at multiple sites
produced data in time order but not synchronously.  Simulation and
processing did not preserve the same ``event'' multiplicity.  Multiple
sources of events (many independent backgrounds in addition to signal)
must be properly mixed in time and at multiple stages in the
processing chain.  Finally, delayed coincidence in time within one
detector stream and between those of different detectors had to be
formed.  The flexibility of Gaudi allowed Daya Bay to extend its very
event processing model to add the support necessary for these
features.

\subsection{CMSSW and art}

In 2005, the CMS Experiment developed their current software framework, CMSSW \cite{cmssw}, as a replacement to the previous ORCA framework. The framework was built around two guiding principles: the modularity of software development and that exchange of information between modules can only take place through data products. Since implementing the CMSSW, the complexity of the CMS reconstruction software was greatly reduced compared with ORCA and the modularity lowered the barrier to entry for beginning software developers.\footnote{Thanks to Dr. Liz Sexton-Kennedy and Dr. Oli Gutsche for useful discussions concerning the history and design of CMSSW.}

The CMS Software Framework is designed around four basic elements: the framework, the event data model, software modules written by physicists, and the services needed by those modules\cite{cmssw_web}. The framework is intended to be a lightweight executable (cmsRun) that loads modules dynamically at run time. The configuration file for cmsRun defines the modules that are part of the processing and thus the loading of shared object libraries containing definitions of the modules. It also defines the configuration of modules parameters, the order of modules, filters, the data to be processed, and the output of each path defined by filters. The event data model (EDM) has several important properties: events are trigger based, the EDM contains only C++ object containers for all raw data and reconstruction objects, and it is directly browsable within ROOT. It should be noticed that the CMSSW framework is not limited to trigger based events, but this is the current implementation for the CMS experiment. Another important feature of the EDM over the ORCA data format was the requirement that all information about an event is contained within a single file. However, file parentage information is also kept so that if object from an input file are dropped (eg. the raw data) that information can be recovered by reading both the parent file and the current file in downstream processes. The framework was also constructed such that the EDM would contain all of the provenance information for all reconstructed objects. Therefore, it would be possible to regenerate and reproduce any processing output from the raw data given the file produced from CMSSW. Another element of the framework that is useful for reproducibility is the strict requirement that no module can maintain state information about the processing, and all such information must be contained within the base framework structures.

The art framework is an event processing framework that is an evolution of the CMSSW framework. In 2010, the Fermilab Scientific Computing Division undertook the development of an experiment-agnostic framework for use by smaller experiments that lacked the personpower to develop a new framework. Working from the general CMSSW framework, most of the design elements were maintained: lightweight framework based on modular development, event data model, and services required for modules. The output file is ROOT browsable and maintains the strict provenance requirements of CMSSW. For Intensity and Cosmic Frontier experiments, the strict definition of an event being trigger based isn't appropriate and so this structuring was removed and each instance of art allows the experiment to define the event period of interest as required. art is currently being used by the Muon g-2, $\mu2e$, NOvA, $\mu BooNE$, and LBNE/ 35T prototype experiments.

CMSSW did initially have some limitations when implemented, the most significant being the use of non-interpreted, run-time configuration files defined by the FHiCL language. The significance of this being that configuration parameters could not be evaluated dynamically and were required to be explicitly set in the input file. This limitation meant it was impossible to include any scripting within the configuration file. This limitation was recognized by the CMS Collaboration and they quickly made the choice to instead transition to Python (in 2006) based configuration files. At that time, a choice was made that the Python evaluation of configuration code would be distinctly delineated from framework and module processing. Therefore, once the configuration file was interpreted, all configuration information was cast as const within C++ objects and immutable. Due to the requirement within CMSSW for strict inclusion of provenance information in the EDM, the dynamic evaluation of configuration files then cast as const parameters and stored in the EDM was not considered a limitation to reproduction from raw data. When the art framework was forked from CMSSW in 2010, the art framework reverted back to using FHiCL language configuration files, and, while acceptable to experiments at the time of adoption, some consider this a serious limitation.

One of the challenges faced by the art framework has been the portability of the framework to platforms other than Scientific Linux Fermilab or Cern. The utilization of the Fermilab UPS and cetbuildtools products within the build and release system that was integrated into the art suite resulted in reliance upon those products that is difficult to remove and therefore port to other platforms (OS X, Ubuntu, etc). The CMSSW framework was implemented for CMS such that the build system was completely available from source and mechanisms for porting to experiment-supported platforms is integrated into the build system. While portability of art is not an inherent problem of the software framework design, and is currently being addressed by both Fermilab SCD and collaborative experiments, it serves as a significant design lesson when moving forward with art or designing other frameworks in the future.

\subsection{IceTray}

IceTray\cite{icetray} is the software framework used by the IceCube experiment and also ported to SeaTray for the Antares experiment. The framework is similar to other previously described frameworks in that it takes advantage of modular design for development and processing. Processing within the framework has both analysis modules and services similar to those described for Gaudi, CMSSW, and art. The IceTray framework and modules are written in the C++ language. The data structure for IceTray is designated a ``frame" and contains information about geometry, calibration, detector status, and physics events. Unlike other frameworks described, IceTray allows for multiple frames to be active in a module at the same time. This was implemented due to the nature of the IceCube detector and the need to delay processing an ``event" until information from more than the current frame is analyzed. This is accomplished through the use of a uniquely designed I/O mechanism utilizing Inboxes and Outboxes for modules. A module can have any number of Inboxes and Outboxes. The development of IceTray was done within the IceCube experiment based upon a specific set of requirements in 2003.

\subsection{Opportunity for improvement}

Some best practices relevant to event processing frameworks are identified:

\begin{description}
\item[open community] Make source-code repositories, bug tickets and
  mailing lists (user and developer) available for anonymous reading
  and lower the barrier for accepting contributions from the community.

\item[modularity] Separate the framework code into modular compilation
  units with clear interfaces which minimize recompilation.  The
  system should work when optional modules are omitted and allow
  different modules to be linked at run-time.

\item[documentation] Produce descriptions of the concepts, design and
  implementation of the framework and guides on installation,
  extension and use of the framework.
\end{description}

\noindent The community should work towards making one event
processing framework which is general purpose enough to service
multiple experiments existing at different scales.  This framework
should be ultimately developed by a core team with representation from
multiple, major stake-holder experiments and with an open
user/developer community that spans other experiments.  Steps to reach
this goal may include:

\begin{itemize}
\item Form an expert working group to identify requirements and
  features needed by such a general use event processing framework.
  Much of this exist in internal and published notes and needs to be
  pulled together and made comprehensive.
\item The working group should evaluate existing frameworks with
  significant user base against these requirements and determine what
  deficiencies exist and the amount of effort required to correct
  them.
\item The working group should recommend one framework, existing or
  novel, to develop as a widely-used, community-supported project.
\item The working group should conclude by gauging interest in the
  community, survey experiments to determine what involvement and use
  can be expected and determine a list of candidate developers for the next
  step.
\item Assemble a core team to provide this development and support
  (something similar to the ROOT model).  Direct support, which is
  independent from specific experiment funding, for some significant
  portion of this effort is recommended.
\end{itemize}

\pagebreak
\section{Software Development}


\subsection{Description}
\label{sec:swdevpartitions}

The tools supporting the full software development life-cycle can be
partitioned into these orthogonal categories.

\begin{description}
\item[Code Repositories] store a historical record of revisions to a
  code base including information on when a change is made, the
  identity of the developer and some note explaining the change.
  Repositories may be organized to hold a single logical unit of
  source code (ie, a \textit{source package}) or may include multiple
  such units relying on some internal demarcation.  They allow
  diverging lines of development, merging these lines and placing
  labels to identify special points in the history (ie, release tags).

\item[Package Build System] contains tools applied to the files of
  \textit{source package} in order to transform them into some number
  of resulting files (executable programs, libraries).  Typically the
  system executes some number of commands (compilers, linkers) while
  applying some number of build parameters (debug/optimized
  compilation, locating dependencies, activating code features).  This
  system may directly install the results of the build to some area or
  in addition it may collect the build results into one or more
  \textit{binary packages}.

\item[Release Configuration] contains tools or specifications for the
  collection of information needed to build a cohesive suite of
  packages.  It includes the list of packages making up the suite,
  their versions, any build parameters, file system layout policy,
  source locations, any local patch files and the collection of
  commands needed to exercise the \textit{package build system}. 

\item[Release Build System] contains tools or processes (instructions)
  that can apply a \textit{release configuration} to each
  \textit{package build system} in the software suite.  This process
  typically iterates on the collection of packages in an order that
  honors their inter-dependencies.  As each package is built, the
  \textit{release build system} assures it is done in a proper context
  containing the build products of dependencies and ideally,
  controlling for any files provided by the general operating system
  or user environment. This system may directly install the results of
  the build to some area and it may collect the build results into one
  or more \textit{binary packages}.

\item[Package Installation System] contains tools that, if they are
  produced, can download and unpack \textit{binary packages} into an
  installation area.  This system is typically tightly coupled to the
  binary package format.  It may rely on meta data internal or
  external to the binary package file in order to properly resolve
  dependencies, conflicts or perform pre- and post-installation
  procedures.  The system may require privileged access and a single
  rooted file system tree or may be run as an unprivileged user and
  allow for multiple and even interwoven file system trees.

\item[User Environment Management] contains tools that aggregate a
  subset of installed software in such a way that the end user may
  properly execute the programs it provides.  This aggregation is
  typically done through the setting of environment variables
  interpreted by the shell, such as \texttt{PATH}.  In other cases the
  bulk of aggregation is done via the file system by copying or
  linking files from some installation store into a more localized
  area and then defining some minimal set of environment variables.
  In the case where software is installed as system packages 
  environment management may not be required.

\item[Development Environment Management] contains tools to assist the
  developer in modifying existing software or writing novel packages.
  Such tools are not strictly required as a developer may use tools
  from the above categories to produce a personal release.  However,
  in practice this makes the development cycle (modify-build-test
  loop) unacceptably long.  To reduce this time and effort, existing
  release builds can be leveraged, installation steps can be minimized
  or removed, and environment management can be such as to use the
  build products in-place.  Care is needed in designing such tools to
  mitigate interference between individual developers while allowing
  them to synchronize their development as needed.

\item[Continuous Integration] contains tools and methodologies for
  developing and exercising the code in order to validate changes,
  find and fix problems quickly, and vet releases.

\item[Issues Tracker] contains tools to manage reporting,
  understanding and addressing problems with the software, requests
  for new features, organizing and documenting releases.

\end{description}

The following sections give commentary on what aspects are successful
for providing general, cross-experiment benefit and what failings are
identified.  Explicit examples and areas where improvement may be made are given.

\subsection{Follow Free Software}

The Free Software (FS) and the Open Source (OS) communities have a
large overlap with HEP in terms of how they develop and use software.
FS/OS has been very successful in achieving beneficial sharing of
software, of course, largely due to that being a primary goal of the
community.  It is natural then for the HEP software community to try
to emulate FS/OS.

Of course, HEP does already benefit greatly from adopting many
libraries and tools from FS/OS.  The community is relatively open with
its software development (in comparison to, for example, industry).  

There are however some ways in which HEP community currently differs
from the FS/OS.  Some are necessary and some are areas where
improvements can be made.

\begin{itemize}
\item Physics is the primary focus, not software.  Of course this is
  proper.  But, software is often not considered as important as other
  secondary aspects such as detector hardware design despite the fact
  that detector data is essentially useless today without quality software.  Even
  in areas where software is the focus, often the ``hard core''
  software issues are down-played or considered unimportant.
\item The use and development of HEP software is often tightly
  intertwined.  End users of software are often its developers.
  Making formal releases is often seen as a hindrance or not 
  performed due to lack of familiarity or access to easily usable
  release tools.
\item HEP software typically must be installed with no special
  permissions (non-``root''), in non-system locations, and with
  multiple versions of the software available on the same system.
  User/developers will often need to maintain locally modified copies
  of the software that override but otherwise rely on some centrally
  shared installation.
\item Versions matter a lot until they don't.  A single code commit
  may radically change results and so upgrades must be done with care
  and changes constantly validated.  Old versions must be kept accessible
  until new ones are vetted.  They then become unimportant but must be
  forever reproducible in case some issue is found in the future which
  requires rerunning of the old version.
\item HEP software suites tend to be relatively large, often with the
  majority consisting of software authored by HEP physicists.  Their
  design often requires an all-or-nothing adoption.  Lack of careful
  modular components with well defined interfaces lead to design
  complexity and practical problems such as compilation time.
  Dependencies must be carefully handled and tested when lower-layer
  libraries are modified.
\end{itemize}

\subsection{Category Integration}

The categories described in section \ref{sec:swdevpartitions} present
some ideal partitioning.  Real world tools often cover multiple
categories.  When this integration is done well it can be beneficial.
Where it is not done well it can lead to lock-in, lack of portability,
increased maintenance costs and other pathologies.

The functions of Configuration Management Tools (CMT) spans most of these categories.
Its design is such that it provides beneficial integration with some
capability to select the categories in which to apply it.  For
example, it provides a package build system but one which is flexible
enough to either directly execute build commands or to delegate to
another package build system.  This allows building and use of
external packages to achieve symmetry with packages developed by the
experiment.  The configuration system is flexible enough to tightly
control versions for releases or to relax dependency conditions
suitable for a development context.  The same information used to
build packages is used to generate shell commands to configure
end-user environment.

CMT was initially used by LHC experiments but has successfully be
adopted to others outside of CERN (Daya Bay, Minerva, Fermi/GLAST, and
others).  It is used across the three major computer platforms (Linux,
Mac OS X and Windows).

In contrast is the UPS/CET system from Fermilab currently used to
build the art framework and its applications.  UPS itself shares some
passing familiarity to CMT although its implementation is such that
even its proponents do not typically use it fully as it was designed.  Its
entire ability to build packages is largely avoided.  Its other
primary purpose of managing user environment is often augmented with
custom shell scripts.  

The CET portion adds a package build system based on CMake but with
hard wired entanglements with UPS.  It tightly locks in to the source
code which versions of dependencies must be built against and the
mechanism to locate them.  Developers commonly have their own effort
derailed if they attempt to incorporate work from others as any
intervening release forces their development base to become broken and
require reinitializing.  Attempting to port the software (art and
LArSoft) entangled with this build system from the only supported
Linux distribution (Scientific Linux) to another (Debian) was found to
be effectively impossible in any reasonable time frame.  This has led
to an effort by the LBNE collaboration to fully remove the UPS/CET
package build system from this software and replace it with one still
based on CMake but which follows standard forms and best practices.
It took far less effort to reimplement a build system than to adopt
the initial one.  Effort is ongoing to incorporate these changes back
into the original software.

The astrophysics experiment LSST has developed a system, EUPS \cite{lssteups}, 
based on UPS, for code builds which allows local builds on an experiment 
collaborators' laptop or server and which probes the users local machine 
for already installed standard packages (such as python).   This system
may be worth a look for smaller scale experiments \cite{lsstwiki}.

\subsection{Distributed Software Tools}

Network technology has lead to paradigm shifts in binary code
distribution (eg CVMFS) and in distributing data (eg XRootD).  HEP
software development has always been very distributed and it is
important to continue to embrace this.

One successful embrace has been the move to \texttt{git} for managing source
code revisions.  In comparison, any code development that is still
kept in Concurrent Versions System (CVS) or Subversion (SVN) is at a
relative disadvantage in terms of the ability to distribute
development effort and share its results.

Aggregating \texttt{git} repositories along with associated issue trackers, web
content (wikis) to provide a definitive, if only by convention, center
of development is also important.  Some institutions provide these
aggregation services (Fermilab's Redmine) but the full benefit comes
when the software is exposed in a more global way such as through
online repository aggregators like GitHub or BitBucket.

Building software is an area that would benefit from a more
distributed approach.  The traditional model is that the software
needed by the experiment is built from source by a site administrator
or an individual.  In some cases, an institution will take on the job
of building software for multiple experiments such as is done for some
experiments centered at CERN and Fermilab.  While this service is
helpful for users of the platforms supported by the institution, it
tends to lock out users who do not use the officially supported
computer platforms.  These unsupported platforms are otherwise
suitable for use and are often more advanced that then officially
supported ones.  Small incompatibilities build up in the code base
because they go undetected in the relative monoculture created by
limiting support to a small number of platforms.

Distributed software build and installation systems are becoming
established and should be evaluated for adoption.  Examples include
the package management systems found in the Nix and Guix operating
systems.  These allows one individual to build a package in such a way
that it may be used by anyone else.  They also provide many innovative
methods for end-user package aggregation which leverage the file
system instead of polluting the user's environment variables.

Another example is Conda which provides a method to bundle up the
build configuration and a one-package unit of a release build system.
It also provides an end-user tool to install the packaged results.  A
coupled project is Binstar which can be thought of as a mix between
GitHub and the Python Package Index (PyPI).  It allows upload and
distribution of packages built by Conda for later end-user download
and installation.

HEP community software projects and individual experiments can make
use of either the Nix/Guix or Conda/Binstar approaches to provide
ready-to-use code binaries to any networked computer in a trusted
manner.
Sharing and coordinating the production of these packages would take
additional effort but this will be paid back by the reduction of so
much redundant effort that goes into building the same package by each
individual, experiment or project.

\subsection{Automate}

The majority of most HEP software suites are composed of four layers:
the experiment software on top is supported by general-use HEP
software.  Below that is FS/OS packages which may be included in some
operating system distributions but in order to control versions and to
provide a uniform base for those OS distributions which do not provide
them, they are build from source.  Finally, there is some lowest layer
provided by the OS.  Each experiment draws each of these lines
differently and some may choose to blur them.

To produce proper and consistent release configurations and to track
them through time is challenging.  Once created, in principle, a
system can then apply these configurations in a way that automates the
production of the release build.  Without this automation the amount
of effort balloons.  This is only partially mitigated by distributing
the build results (addressed above).

Some experiments have developed their own build automation based on
scripts.  These help the collaborators but they are not generally
useful.

CERN developed LCGCMT which, in part, provides automated building of
``externals'' via the \verb|LCG_Builders| component.  This system is
specifically tied to CMT and is particularly helpful if CMT is adopted
as a release build system.  This mechanism has been adopted by groups
outside of CERN, specifically those that also adopted the Gaudi event
processing framework.  It has been specifically adopted by other
experiments.

Growing out of the experience with custom experiment-specific
automation and LCGCMT, the Worch~\cite{worch} project was developed to provide
build ``orchestration''.  This includes a release configuration method
and an automated release build tool.  It is extensible to provide
support for the other software development tool categories.  For
example, it has support for producing needed configuration files to
provide support for using Environment Modules as a method for end-user
environment management.

\subsection{Opportunity for improvement}

Some best practices in the area of software development tools are:

\begin{description}
\item[Leverage Free Software] Rely on Free Software / Open Source and
  do not become shackled to proprietary software or systems.
\item[Portability] Do not limit development to specific platform or
  institutional details.
\item[Automate] Produce builds and tests of software stacks in an
  automated manner that is useful to both end-user/installers and
  developers.
\end{description}

\noindent Some concrete work that would generally benefit software development efforts in HEP include:

\begin{itemize}
\item Form a cross-experiment group to determine requirements for
  tools for build automation, software release management, binary
  packaging (including their format), end-user and developer
  environment management.
\item Form teams to design, identify or implement tools meeting these
  requirements.
\item Assist experiments in the adoption of these tools.
\end{itemize}

\pagebreak
\section{Data Management}
\label{data}


\subsection{Definition}
\textbf{Data Management} is any \textit{content neutral} interaction with the data, e.g. it is the \textit{data flow}
component of the larger domain of  \textbf{Workflow Management} (see \ref{workflow_workload}). It addresses issues of data storage
and archival, mechanisms of data access and distribution and  curation -- over the full life cycle of the data. In order to remain within
the scope of this document we will
concentrate on issues related to data distribution, metadata and catalogs, and won't cover issues of mass storage
in much detail (which will be covered by the \textit{Systems} working group). Likewise, for the most part network-specific issues fall outside of our purview.


\subsection{Moving Data}
\subsubsection{Modes of Data Transfer and Access}
\label{data_xfer}
In any distributed environment (and most HEP experiments are prime examples of that) the data are typically stored at multiple locations,
for a variety of reasons, and over their lifetime undergo a series of transmissions over networks, replications and/or deletions, with attendant bookkeeping
in appropriate catalogs. Data networks utilized in research can span \textit{hundreds} of Grid sites across multiple continents.

In HEP, we observe a few different and distinct modes of moving and accessing data (which, however, can be used in a complementary fashion).
Consider the following:

\begin{description}
\item[Bulk Transfer] In this conceptually simple case, data transport from point A to point B is automated and augmented
with redundancy and verification mechanism so as to minimize chances of data loss. Such implementation may be needed,
for example, to transport ``precious'' data from the detector to the point of permanent storage.
Examples of this can be found in SPADE (data transfer system used in Daya Bay) and components of SAM~\cite{SAM} and File Transfer Service at FNAL.
Similar functionality (as a part of a wider set) is implemented in the \textit{Globus Online} middleware kit~\cite{globus}.

\item[Managed Replication] In many cases the data management strategy involves creating replicas of certain segments of the data (datasets, blocks, etc)
at participating Grid sites. Such distribution is done according to a previously developed policy which may be based on storage capacities of 
the sites, specific processing plans (cf. the concept of \textit{subscription}), resource quota and any number of other factors. Good examples of this type of systems are found in
ATLAS (\textit{Rucio)} and CMS (\textit{PhEDEx}), among other experiments~\cite{rucio_chep13,phedex_chep09}.

\item[``Data in the Grid'' (or Cloud)] In addition to processing data which is local to the processing element (i.e. local storage), such as a Worker Node
on the Grid, it is possible to access data over the network, provided there exists enough bandwidth between the remote storage
facility or device, and the processing element. There are many ways to achieve this. Examples:
\begin{itemize}
\item using \textit{http} to pull data from a remote location before executing the payload job. This can involve private data servers or public cloud facilities.
\item utilizing \xrootd~\cite{xrootd,xrootd_web} over WAN to federate storage resources and locate and deliver files transparently in a ``just-in-time'' manner.
\item sharing data using middleware like Globus~\cite{globus}.
\end{itemize}

\end{description}

Choosing the right approaches and technologies is a two-tiered process. First, one needs to identify the most
 relevant use cases and match them to categories such as outlined above (e.g. replication vs network data on demand). Second, within
 the chosen scenario, proper solutions must be identified (and hopefully reused rather than reimplemented).

\subsubsection{From MONARC to a Flat Universe}
The \textbf{MONARC} architecture is a useful case study, in part because it was used in the LHC Run 1 data processing campaign,
and also because it motivated the evolution of approaches to data management which is currently under way.
It stands for \textit{Models of Networked Analysis at Regional Centers} \cite{monarc}.
At the heart of  MONARC  is a manifestly hierarchical organization of computing centers in terms of
data flow, storage and distribution policies defined based on characteristics and goals of participating sites. The sites
are classified into ``Tiers'' according to the scale of their respective resources and planned functionality, with ``Tier-0'' denomination
reserved for central facilities at CERN, ``Tier-1'' corresponding to major regional centers while ``Tier-2'' describes
sites of smaller scale, to be configured and used mainly for analysis of the data (they are also used
to handle a large fraction of the Monte Carlo simulations workload). Smaller installations and what is termed ``non-pledged resources'' 
belong to Tier-3 in this scheme, implying a more \textit{ad hoc} approach to data distribution and handling of the computational
workload on these sites. The topology of the data flow among the
Tiers can be described as somewhat similar to a Directed Acyclic Graph (DAG), where data undergoes processing steps 
and is distributed from Tier-0 to a number of Tier-1 facilities, and then on to Tier-2 sites -- but Tiers of the same rank
do not share data on ``p2p'' basis.
This architecture depends on coordinated operation of two major components:
\begin{itemize}
	\item The Data Management System, that includes databases necessary to maintain records of the data parameters and location,
	and which is equipped with automated tools to move data between 
	computing centers according to chosen data processing and analysis strategies and algorithms. 
	An important component of the data handling is a subsystem for managing \textit{Metadata}, i.e. information
	derived from the actual data which is used to locate specific data segments for processing based on 
	certain selection criteria.
	
	\item The Workload Management System (WMS) -- see Section~\ref{wms} -- which distributes computational payload in accordance  
	with optimal resource availability and various applicable policies. It typically also takes
	into account data locality in order to minimize network traffic and expedite execution. A mature 
	and robust WMS also contains efficient and user-friendly monitoring capabilities, which allows 
	its operators to monitor and troubleshoot workflows executed on the Grid.
	
\end{itemize}

While there were a variety of factors which motivated this architecture, considerations of overall efficiency, given
limits of storage capacity and network throughput, were the primary drivers in the MONARC model. In particular,
reconstruction, reprocessing and some initial stages of data reduction are typically done at the sites with
ample  storage capacity so as to avoid moving large amount of data over the network. As such, it can be argued 
that the MONARC architecture was ultimately influenced by certain assumptions about bandwidth, performance 
and reliability of networks which some authors now call ``pessimistic'' \cite [p.~105]{lhc_model_update}.

At the time when LHC computing was becoming mature, great progress was made in improving
characteristics of the networks serving the LHC projects. New generation of networks has lower
latencies, lower cost per unit of bandwidth and higher capacity. This applies to both local and wide 
area networks  \cite[p.104]{lhc_model_update}. This development opens new and significant possibilities 
which were not available until relatively recently; as stated in \cite{lhc_model_update},

\begin{quote}
The performance of the network has allowed a more flexible model in terms of data access:
	
	\begin{itemize}
		\item Removal of the strict hierarchy of data moving down the tiers, and allowing a
		more peer-2-peer data access policy (a site can obtain data from more or less any 
		other site);
		
		\item The introduction of the ability to have remote access to data, either in obtaining
		missing files needed by a job from over the WAN, or in some cases actually
		streaming data remotely to a job.
		
	\end{itemize}
\end{quote}

In practice, this new model results in a structure which is more ``flat'' and less hierarchical \cite{lhc_model_update}, \cite{courier_update} -- and indeed
resembles the ``p2p'' architecture.

In principle, this updated architecture does not necessarily require new networking and data transmission 
technologies when compared to MONARC, as it mainly represents a different logic and policies for 
distribution of, and access to data across multiple Grid sites. Still, there are a number of differing 
protocols and systems which are more conducive to implementing this approach than others, for a variety of reasons:

\begin{itemize}
	\item Reliance on proven, widely available and low-maintenance tools to actuate data transfer (e.g. utilizing HTTP/WebDAV).
	\item Automation of data discovery in the distributed storage.
	\item Transparent and automated "pull" of required data to local storage.
\end{itemize}

One outstanding example of leveraging the improved networking technology is XRootD - a system which facilitates \textit{federation} of widely 
distributed resources~\cite{xrootd_fed,xrootd_snowmass}. While its use in HEP is widespread, two large-scale applications deserve a special mention: 
it is employed  in the form of CMS's ``Any Data, Anytime, Anywhere'' (AAA)
project and ATLAS's ``Federating ATLAS storage systems using Xrootd" (FAX) project, both of which rely
on XRootD as their underlying technology. ``These systems are already giving experiments and
individual users greater flexibility in how and where they run their workflows by making data more globally
available for access. Potential issues with bandwidth can be solved through optimization and prioritization''\cite{xrootd_snowmass}.

\subsection{Metadata, Data Catalogs and Levels of Data Aggregation}
To be able to locate, navigate and manage the data it has to be \textit{described}, or characterized. Metadata (data derived from the data) is
therefore a necessary part of data management. The  list of possible types of metadata is long. Some key ones are:

\begin{itemize}
\item Data Provenance: for raw data, this may include information on when and where it was taken. For processed data,
it may specify which raw data were used. For many kinds of data, it is important to track information about calibrations used, etc.

\item Parentage and Production Information: one must keep track of software releases and its configuration details in each production step,
be able to trace a piece of data to its origin (e.g. where it was produced, by which Task ID etc), etc.


\item Physics: this may include analysis summary information or a specific feature characterizing a segment of data, e.g. type of events selected, from which trigger stream data was derived, detector configuration.
\item Physical information: this might include the file size, check sum, file name, location, format, etc.
\end{itemize}
Generally speaking, a data catalog combines a file catalog i.e. information about where the data files are stored,
with additional metadata that may contain a number of attributes (physics, provenance etc). This enables the construction of logical (virtual)
data sets like 'WIMPcandidatesLoose' and makes it possible for users to select a subset of the available data, and/or ``discover'' the presence and locality of data which
is of interest to the user.
Grouping of data into datasets and even larger aggregation units helps handle complexity of processing which involved a very large number of induvudual files.
Here are some examples:

\begin{description}
\item[Fermi Data Catalog:] Metadata can be created when a file is registered in the database. A slightly
different approach was chosen by the Fermi Space Telescope Data Catalog. In addition to the initial metadata, it has a
data crawler that would go through all registered files and extract metadata like number of events etc. The advantage is
that the set of metadata then can be easily expanded after the fact - you just let loose the crawler with the list
of new quantities to extract which then is just added to the existing list of metadata). Obviously this only works for 
metadata included in the file and file headers.  Note that since the Fermi Data Catalog is independent of any
workflow management system, any data processing metadata will have to be explicitly added.

\item[SAM:] (Sequential Access Model) is a data handling system
  developed at Fermilab.  It is designed to track locations of files
  and other file metadata.  A small portion of this metadata schema is
  reserved for SAM use and experiments may extend it in order to store
  their quantities associated with any given file.  SAM allows for the
  defining of a \textit{dataset} as a query on this file metadata.
  These datasets are then short hand which can then be expanded to
  provide input data to for experiment processes.  Beyond this role as
  a file catalog SAM has additional functionality.  It can manage the
  storage of files and it can serve an extended roll as part of a workflow
  management system.  It does this through a concept called
  \textit{projects} which are managed processes that may deliver files
  to SAM for storage management and deliver files from storage elements
  to managed processes. SAM maintains state information for files in
  active projects to determine which files have been processed, which process
  analyzed each file, and files consumed by failed processes. The installation footprint required
  for SAM to be used at a participating site depends on the
  functionality required.  Lightweight access to catalog functionality
  is provided via the SAM Web Services component through a REST web
  interface which includes a Python client module.  Full features,
  including file management, requires a SAM \textit{station}
  installation and these exist at a small number of locations.

\item[ATLAS:] Distributed Data Management in ATLAS (often termed \textit{DDM}) has always been one of its focus areas, in part due to sheer volume of
data being stored, shared and distributed worldwide (on multi-petabyte scale), and to the importance of optimal data placement to ensure efficiency
and high throughput of processing~\cite{atlas_ddm_chep12}. Just like with other major components of its systems, ATLAS has evolved its
data management infrastructure over the years. The systems currently utilized is Rucio~\cite{rucio_chep13}. We shall briefly consider basic
concepts and entities in this system pertaining to this section.

The atomic unit of data in ATLAS is file. Above that, there are levels of data aggregations, such as:
\begin{itemize}
\item \textit{dataset} Datasets are the operational unit of replication for DDM, i.e., they may be transferred to grid sites, whereas single files
may not. Datasets in DDM may contain files that are in other datasets, i.e., datasets may overlap.
\item \textit{container} Container is a collection of datasets. Containers are not units of replication, but allow
large blocks of data, which could not be replicated to single site, to be described in the system.
\end{itemize}

There are a few categories of metadata in Rucio:
\begin{itemize}
\item  System-defined attributes (e.g. size, checksum etc)
\item Physics attributes (such as number of events)
\item Production attributes (parentage)
\item Data management attributes
\end{itemize}

\item[CMS:] CMS also employs the concept of a dataset. Metadata resides in, and is being handled by the ``The Data Bookkeeping Service'' (DBS).
This service maintains information regarding the provenance, parentage, physics attributes and other type of metadata necessary for efficient processing.
The Data Aggregation Service (DAS)  is another critical component of the CMS Data Management System. It ``provides the ability to query CMS data-services via
a uniform query language without worrying about security policies and differences in underlying data representations''~\cite{phedex_chep13}.

\end{description}

\subsection{Small and Medium Scale Experiments}
Small and medium scale research programs often have smaller needs compared to the LHC or other large experiments. In these cases, it won't
be economical or feasible to deploy and operate same kind of middleware on the scale described in previous sections.
Data is often be stored in a single or just a few geographical locations ('host laboratories'), and
data processing itself is less distributed. However, many experiments today have data (or will have data) characterized by volumes
and complexity large enough to create and demand a real data management system instead of resorting to manual mode (files in unix directories and wiki pages). 
In fact, we already find that some of the same elements, i.e. extensive metadata, data catalogs, XRootD, are also used by some smaller 
experiments. The main challenge here is the very limited technical expertise and manpower available to develop, adapt and 
operate this sort of tools.

With Cloud technology recently becoming more affordable, available and transparent for use in a variety of applications, smaller
scale collaborations are making use of services such as Globus~\cite{globus} to perform automated managed data transfers (cf.~\ref{data_xfer}),
implement data sharing and realize the ``Data in the Cloud'' approach. For small and mid-scale projects, platforms like \textit{Google Drive}
and \textit{Dropbox} offer attractive possibilities to share and store data at a reasonable cost, without having to own much of storage and
networking equipment and to deploy a complex middleware stack.

\subsection{Opportunities For Improvement}
\subsubsection{Modularity}
One problem with Data Management systems is that they often tend to become monolithic as more and more functionality is 
added (organically) -- see Sec.~\ref{3domains}. While this may make it easier to operate in the short term, it makes it more difficult to maintain over 
the long term. In particular, it makes it difficult to react to technical developments and update parts of the system. It's 
therefore critical to make the system as modular as possible and avoid tight couplings to either specific technologies or 
to other parts of the ecosystem, in particular the coupling to the Workload Management System. Modularity should therefore be 
part of the core design and specifically separating the Metadata Catalogs from Data Movement tools, with carefully designed
object models and APIs. This would also make reuse easier to achieve.

\subsubsection{Smaller Projects}
Smaller experiments have different problems. Most small experiments have or will enter the PB era and can no longer use a manual 
data management system built and operated by an army of grad students. They need modern data management tools. However, they have neither the 
expertise to adapt LHC scale tools for their use, neither the technical manpower to operate them. Simplifying and downscaling existing 
large scale tools to minimize necessary technical expertise and manpower to operate them, even at the price of decreasing functionality, 
may therefore be a good option.

A second option is to take existing medium-scale data handling tools and repackage them for more general use. The problem is, 
however, somewhat similar to what is described above. Often these systems have become monolithic, have strong couplings to certain 
technologies and significant work may be necessary to make them modular. This can be difficult to achieve within the limited resources
available and will need dedicated support.

Finally, a few recent Cloud solutions became available (and are already used by small to medium size project), such as Globus~\cite{globus},
Google Drive and Dropbox, among others. They do provide a lot of necessary functionality for data distribution and sharing, and perhaps
provide an optimal solution at this scale, when combined with a flexible and reusable Metadata system (see notes on modularity above).

\subsubsection{Federation}
Last, the success of Federated Storage built on XRootD shows the importance of good building blocks and how they can be arranged into 
larger successful systems.



\pagebreak
\section{Workflow and Workload Management}
\label{wms}

\subsection{The Challenge of the Three Domains}
\label{3domains}
In the past three decades, technological revolution in industry has enabled and was paralleled by the growing complexity in the field of scientific computing, where 
more and more sophisicated methods of data processing and analysis were constantly introduced, oftentimes at a dramatically increased scale.
Processing power and storage were becoming increasingly decentralized, leading to the need to manage these distributed resources in an optimal manner.
On the other hand, increasing sophistication of scientific workflows created the need to support these workflows in the new distributed computing medium.
Rapid evolution of the field and time pressures to deliver in this competitive environment led to design and implementation of complete (to varying degrees)
and successful solutions to satisfy the needs of specific  research projects. In general, this had two consequences:

\begin{itemize}
\item Integrated and oftentimes -- not always -- project-specific design of workflow and workload management (see~\ref{workflow_workload} for definitions).
\item Tight coupling of workflow and workload management to data handling components.
\end{itemize}

We observe that there are essentially \textbf{three interconnected domains} involved in this subject: Workflow Management, Workload Management,
and Data Management. In many systems (cf. Pegasus~\cite{pegasus}) some of these domains can be ``fused'' (e.g. Workflow+Workload).
In the following, we  bring together a few standard definitions and real life examples to help clarify
relationships among these domains and in doing so form the basis for possible HEP-FCE recommendations. Our goal will be twofold:
\begin{itemize}
\item to identify the features and design considerations proven to be successful and which can serve as guidance going forward.
\item to identify common design and implementation elements and to develop understanding of how to enhance \textit{reusability} of existing and future systems of this kind.
\end{itemize}

\subsection{Description}

\subsubsection{Grid and Cloud Computing}
\label{cloud}
According to a common definition, \textbf{Grid Computing} is the collection of computer resources from multiple locations to reach a common goal. Oftentimes additional
characteristics added to this include decentralized administration and management and adherence to open standards. It was formulated as a concept in early 1990s,
and motivated by the fact that computational tasks handled by large research projects reached the limits of scalability of most individual computing sites.
On the other hand, due to variations in demand, some resources were underutilized at times. There is therefore a benefit in implementing a federation of computing
resources, whereby large spikes in demand can be handled by federated sites, while ``backfilling'' the available capacity with lower priority tasks submitted by a
larger community of users. Technologies developed in the framework of Grid Computing (such as a few reliable and popular types of \textit{Grid Middleware})
became a major enabling factor for many scientific collaborations including nuclear and high-energy physics.

\textbf{Cloud Computing} is essentially an evolution of the Grid Computing concept, with implied higher degree of computing resources and data
storage abstraction, connectivity and transparency of access. In addition, Cloud Computing is characterized by widespread adoption of \textit{virtualization} - which is
also used in the traditional Grid environment but on a somewhat smaller scale. At the time of writing, ``Cloud'' prominently figures in the context of commercial services available on
the Internet, whereby computing resources can be ``rented'' for a fee in the form of a Virtual Machine allocated to the user, or a  number of nodes in the Cloud can
be dynamically assigned to perform a neccesary computational task - often as a transient, ephemeral resource. Such dynamic, on-demand characteristic of the
Cloud led to it being described as an ``elastic'' resource. This attribute is prominently featured in the name of the ``Amazon Elastic Compute Cloud'' -- ``the EC2''.
This is an example of a \textit{public} Cloud, available to most entities in the open marketplace. Some organizations choose to deploy Cloud technology on the
computing resources directly owned and controlled by them, which is referred to as \textit{private} Cloud.

Regardless of the obvious differentiation of Cloud computing (due to its characteristics as \textit{a utility computing platform} and pervasive reliance on virtualization),
many of its fundamental concepts and challenges are common with its predesessor, Grid Computing. In fact, the boundary is blurred even further by existing efforts to
enhance Grid middleware with tools based on virtualization and Cloud API which essentially extend ``traditional'' Grid resources with on-demand, elastic Cloud
capability~\cite{star_acat11}, leading to what is essentially a hybrid system. The two are often seen as ``complementary technologies that will coexist at
different levels of resource abstraction'' \cite{atlas_cloud_chep13}. Moreover,
\begin{quote}
\textit{In parallel, many existing grid sites have begun internal evaluations of cloud technologies (such as Open Nebula or OpenStack) to reorganize the internal management of their computing resources.}
(See \cite{atlas_cloud_chep12})
\end{quote}

In recent years, there are a few open-source, community developed and supported Cloud Computing platforms that reached maturity, such as OpenStack~\cite{openstack}.
It includes a comprehensive set of components such as ``Compute'', ``Networking'', ``Storage'' and others and is designed to run on standard commodity hardware.
It is deployed at scale by large organizations and serves as foundation for commercial Cloud services such as HP Helion~\cite{helion}. This technology allows pooling
of resources of both public and private Clouds, resulting in the so-called \textit{Hybrid Cloud}, where technology aims to achieve the best characteristics of both private
and public clouds.

There are no conceptual or architectural barriers for running HEP-type workflows in the Cloud, and in fact, efforts are well under way to implement this
approach~\cite{atlas_cloud_chep12}. However, there are caveats such as
\begin{itemize}
\item Careful overall cost analysis needs to be performed before each decision to deploy on the Cloud, as it is not universally cheaper than
resources already deployed at research centers such as National Laboratories. At the same time, the Cloud is an efficient way to handle peak demand
for the computing power due to its elasticity. It should also be noted that some national supercomputing centers have made part of their 
capacity available for more general high throughput use and may be an cost effective alternative.
\item Available bandwidth for staging in/staging our data in the Cloud (and again, its cost) need to be quantified and gauged against the project requirements.
\item Cloud storage cost may be an issue for experiments handling massive amounts of data~\cite[p.~11]{atlas_cloud_chep12}
\end{itemize}

\subsubsection{From the Grid to Workload Management}
\label{from_grid_to_workload}
Utilization of the Grid sites via appropriate middelware does establish a degree of resource federation, but it leaves it up to the user to manage data movement and job submission to multiple sites,
track job status, handle failures and error conditions, aggregate bookkeeping information and perform many other tasks. In absence of automation, this does not scale very well and limits the efficacy
of the overall system.

It was therefore inevitable that soon after the advent of reliable Grid middleware multiple physics experiments and other projects started developing and deploying \textbf{Workload Management Systems (WMS)}.
 According to one definition,
\textit{``the purpose of the Workload Manager Service (WMS) is to accept requests for job submission and management coming from its clients and take the appropriate actions to satisfy them''}~\cite{egee_user_guide}.
Thus, one of the principal functions of a Workload Management System can be described as ``brokerage'', in the sense that it matches resource requests to the actual distributed resources
on multiple sites. This matching process can include a variety of factors such as access rules for users and groups, priorities set in the system, or even data locality - which is in fact an important and interesting part of this process~\cite{panda_chep10}.

In practice, despite differing approaches and features, most existing WMS appear to share certain primary goals, and provide solutions to achieve these (to a varying degree). Some examples are:

\begin{itemize}

\item Insulation of the user from the complex and potentially heterogeneous environment of the Grid, and shielding the user from common failure modes of the Grid infrastructure.

\item Rationalization, bookkeeping and automation of software provisioning - for example, distribution of binaries and configuration files to multiple sites.

\item Facilitation of basic monitoring functions, e.g. providing efficient ways to determine the status of jobs being submitted and executed.

\item Prioritization and load balancing across the computing sites.

\item Implementation of interfaces to external data management systems, or actual data movement and monitoring functionality built into certain components of the WMS.

\end{itemize}

We shall present examples of existing WMS in one of the following sections.

\subsubsection{Workflow vs Workload}
\label{workflow_workload}
A \textit{scientific workflow} system is a specialized case of a \textbf{workflow management system}, in which computations and/or transformations and exchange of data are performed according to a defined set of rules
in order to achieve an overall goal ~\cite{grid_workflow_taxonomy,grid_workflow_fit,pegasus}. In the context of this document, this process involves execution on distributed resources. Since the process is
typically largely (or completely) automated, it is often described as ``orchestration'' of execution of multiple interdependent tasks. Workflow systems are sometimes described using the concepts of a \textit{control flow},
which refers to the logic of execution, and \textit{data flow}, which concerns itself with the logic and rules of transmitting data. There are various patterns identified in both control and data flow~\cite{workflow_patterns}.
A complete discussion of this subject is beyond the scope of this document.

A simple and rather typical example of a workflow is often found in Monte Carlo simulation studies performed in High Energy Physics and related fields, where there is a chain of processing steps similar to the pattern below:
\\
\\
\textit{Event  Generation $\Longrightarrow$ Simulation $\Longrightarrow$ Digitization $\Longrightarrow$ Reconstruction}
\\
\\
Patterns like this one may also include optional additional steps (implied or made explicit in the logic of the workflow) such as merging of units of data (e.g. files) for more efficient storage and transmission.
Even the most trivial cases of processing, with one step only, may involve multiple files in input and/or output streams, which creates the need to manage this as a workflow. Oftentimes, however,
workflows that need to be created by researchers are quite complex. At extreme scale, understanding the behavior of scientific workflows becomes a challenge and an object of studies in its own right~\cite{panorama}.

Many (but not all) workflows important in the context of this document can be modeled as
\textbf{Directed Acyclic Graphs} (DAG)~\cite{pegasus,deft1,grid_workflow_taxonomy}.
Conceptually, this level of abstraction of the workflow does not  involve issues of resource provisioning and utilization, monitoring, optimization, recovery from errors, as well as a plethora of other items essential
for efficient execution of workflows in the distributed environment. These tasks are handled in the context of \textit{Workload Management} which we very briefly described in~\ref{from_grid_to_workload}.

In summary, \textbf{we make a distinction between the Workflow Management} domain which concerns itself with controlling the scientific workflow, \textbf{and Workload Management} which
is a domain of resource provisioning, allocation, execution control and monitoring of execution etc. The former is a level of abstraction above Workload Management, whereas the latter is in
turn a layer of abstractions above the distributed execution environment such as the Grid or Cloud.


\subsubsection{HPC vs HTC}
The term \textbf{High-Performance Computing} (HPC) is used in reference to systems of exceptionally high processing capacity (such as individual supercomputers, which are typically highly parallel systems),
in the sense that they handle substantial workload and deliver results over a relatively short period of time. By contrast, HTC involves harnessing to a wider pool
of more conventional resources in order to deliver a considerable amount of computational power, although potentially over longer periods of time. 
\begin{quote}
\textit{``HPC brings enormous amounts of computing power to bear over relatively short periods of time. HTC employs large amounts of computing power for very lengthy periods.''}~\cite{htc}.
\end{quote}

In practice, the term \textit{HTC} does cover most cases of Grid Computing where remote resources are managed for the benefit of the end user and are often made available on prioritized and/or
opportunistic basis (e.g. the so-capped ``spare cycles'' or ``backfilling'', utilizing temporary drops in resource utilization on certain sites to deploy additional workload thus increasing the overall system
throughput). A majority of the computational tasks of the LHC experiments were completed using standard off-the-shelf equipment rather than supercomputers. It is important to note, however,
that modern Workload Management Systems can be adapted to deliver payload to HPC systems such as Leadership Class Facilities in the US, and such efforts are currently under way~\cite{panda_chep13}.

\subsubsection{The Role of Data Management}
In most cases of interest to us, data management plays a crucial role in reaching the scientific goals of an experiment. It is covered in detail separately (see Section~\ref{data}).
As noted above, it represents the \textit{data flow} component of the overall workflow management and therefore needs to be addressed here as well.

In a nutshell, we can distinguish between two different approaches to handling data -- on one hand, managed replication and transfers to sites, and on the other hand, network-centric
methods of access to data repositories such as \xrootd~\cite{xrootd,xrootd_web}.

An important design characteristic which varies widely among Workflow Management Systems is the degree of coupling to the Data Management components. That has significant impact
on reusability of these systems as a more tight coupling usually entails necessity of a larger and more complex stack of software than would otherwise be optimal and has other consequences.

\subsection{Examples}
\label{wms_examples}
\subsubsection{The Scope of this Section}
Workflow and Workload Management, especially taken in conjunction with Data Management (areas with which they are typically interconnected) is a vast
subject and  covering features of each example of WMS in any detail would go well beyond the scope of this document. In the following, we provide references
to those systems which are more immediately relevant to HEP and related fields than others.

\subsubsection{HTCondor and glideinWMS}
\label{htcondor}
HTCondor~\cite{htcondor} is one of the best known and  important set of Grid and High-Throughput Computing (HTC) systems. It provides an array of functionality, such as
a batch system solution for a computing cluster, remote submission to Grid sites (via its \textit{Condor-G} extension) and automated transfer (stage-in and stage-out) of data.
In the past decade, HTCondor was augmented with a  Workload Management System layer, known as \textit{glideinWMS}~\cite{glideinwms}. The latter abstracts remote
resources (Worker Nodes) residing on the Grid and effectively creates a local (in terms of the interface) resource pool accessible by the user.
Putting these resources behind the standard HTCondor interface with its set of utilities is highly beneficial to the users already having familiarity with HTCondor since it geatly shortens the learning curve.
On the other hand, deployment of this system is not always trivial and typically requires a central service to be operated with the desired degree of service level (the so-called ``glidein factory'').

HTCondor has other notable features. One of the most basic parts of its functionality is the ability to transfer data consumed and/or produced by the payload job according to
certain rules set by the user. This works well when used in local cluster situations and is somewhat less reliable when utilized at scale in the context of Grid environment.
One of HTCondor components, \textit{DAGMan}, is a meta-scheduler which uses DAGs (see~\ref{workflow_workload}) to manage workflows~\cite{dagman}. In recent years,
HTCondor was augmented  with Cloud-based methodologies and protocols (cf.~\ref{cloud}).

\subsubsection{Workload Management for LHC Experiments}
This is the list of systems (with a few references to bibliography) utilized by the major LHC experiments - note that in each, we identify components representing layers or subdomains such as Workload Management etc:

\begin{center}
  \begin{tabular}{ c | c | c | c }
    \hline
    \textbf{Project} & \textbf{Workload Mgt} & \textbf{Workflow Mgt} & \textbf{Data Mgt}\\ \hline \hline
    ATLAS & PanDA~\cite{panda_chep10} & ProdSys2 & Rucio~\cite{rucio_chep13}\\ \hline
    CMS  & GlideinWMS~\cite{glideinwms} & Crab3~\cite{crab3_chep12} & PhEDEx~\cite{phedex_chep09,phedex_chep10}\\ \hline
    LHCb  & DIRAC~\cite{dirac_acat09}  & DIRAC Production Mgt & DIRAC DMS\\ \hline
    Alice  & gLite WMS~\cite{glite_chep09} & AliEn~\cite{alien_chep07} & AliEn~\cite{alien_chep07}\\ 
    \hline
  \end{tabular}
\end{center}

\subsubsection{@HOME}
\label{at_home}
There are outstanding examples of open source middleware system for volunteer and grid computing, such as BOINC~\cite{boinc} (the original platform behind SETI@HOME),
FOLDING@HOME and MILKYWAY@HOME \cite{mwathome}. The central part of their design is the server-client architecture, where the clients can be running on a variety of platforms,
such as PCs and game consoles made available to specific projects by volunteering owners. Deployment on a cluster or a farm is also possible.

While this approach to distributed computing won't work well for most experiments of the LHC scale (where moving significant amounts of data presents a perennial problem)
it is clearly of interest to smaller scale projects with more modes I/O requirements. Distributed platforms in this class have been deployed, validated and used at scale.

\subsubsection{European Middleware Initiative}
The European Middleware Initiative~\cite{emi} is a concortium of Grid services providers (such as ARC, dCache, gLite, and UNICORE).
It plays an important role in the the Worldwide LHC Computing Grid (WLCG). The \textit{gLite}~\cite{glite_chep09} middleware toolkit was used by LHC experiments as one of methods
to achieve resource federation on the European Grids.

\subsubsection{Fermi Space Telescope Workflow Engine}
The Fermi workflow engine was originally developed to process data from the Fermi Space Telescope on the SLAC batch farm. The goal was to simplify and 
automate the bookkeeping for tens of thousands of daily batch jobs with complicated dependencies all running on a general use batch farm while 
minimizing the (distributed) manpower needed to operate it. Since it's a general workflow engine it is easily extended to all types of processing 
including 
Monte Carlo simulation and routine science jobs. It has been extended with more batch interfaces and is routinely used to run jobs at IN2P3 Lyon, 
all while being controlled by the central installation at SLAC. It has also been adapted to work in EXO and SuperCDMS.

\subsection{Common Features}

\subsubsection{``Pilots''}
As we mentioned in ~\ref{from_grid_to_workload}, one of the primary functions of a WMS is to insulate the user from the heterogeneous and sometimes complex
Grid environment and certain failure modes inherent in it (e.g. misconfigured sites, transient connectivity problems, ``flaky'' worker nodes etc).
There is a proven solution to these issues, which involved the so called \textbf{late binding} approach  to the deployment of computational payload.

According to this concept, it is not the actual ``payload job'' that is initially dispatched to a  Worker Node residing in a
remote data center, but an intelligent ``wrapper'', sometimes termed a \textbf{``pilot job''}, which first validates the resource, its configuration and
some details of the environment (for example, outbound network connectivity may be tested). In this context, ``binding'' means matching process
whereby the payload job (such as a production job or a user analysis job) which is submitted to the WMS and is awaiting execution is assigned to a
live pilot which has already perfomed validation and configuration of the execution environment (for this reason, this technique is sometimes referred
to as \textbf{``just-in-time workload management''}). Where and how exactly this matching process happens is a subject of a design decision - in PanDA, it's done by
the central server, whereas in DIRAC this process takes place on the Worker Node by utilising the \textit{JobAgent}~\cite{dirac_chep10}.

With proper design,\textbf{ late binding} brings about the following benefits:
\begin{itemize}
\item The part of the overall resource pool that exhibit problems prior to the actual job dispatch is excluded from the matching process by design.
This eliminates a very large fraction of potential failures that the user would otherwise have to deal with and account for, since the resource pool
exposed to the user (or for an automated client performing job submission) is effectively validated.

\item Some very useful diagnostic and logging capability may reside in the pilot. This is very important for troubleshooting and monitoring, which we shall discuss later.
Problematic resources can be identified and flagged at both site and worker node level.

\item In many cases, the overall latency of the system (in the sense of the time between the job submission by the user, and the start of actual execution) will be reduced --
due to the pilot waiting to accept a payload job -- leading to a more optimal user experience (again, cf. the ``just-in-time'' reference).
\end{itemize}

DIRAC was one of the first systems where this concept was proposed and successfully implemented~\cite{dirac_acat09}. This approach also forms the architectural core
of the PanDA WMS~\cite{panda_chep12}.

In distributed systems where the resources are highly opportunistic and/or ephemeral, such as the volunteer computing we mentioned in~\ref{at_home}, this variant
of the client-server model is the essential centerpiece of the design. In BOINC, the ``core client'' (similar to a ``pilot'') performs functions such as maintaining
communications with the server, downloading the payload applications, logging and others~\cite{boinc_client}.

In HTCondor and GlideinWMS (see~\ref{htcondor}) there is no concept of a sophisticated pilot job or core client, but there is a \textit{glidein} agent, which is deployed on Grid resources
and which is a wrapper for the HTCondor daemon process (\textit{startd}). Once the latter is initiated
on the remote worker node it then joins the HTCondor pool. At this point, matching of jobs submitted by the user to HTCondor \textit{slots} becomes possible \cite{glideinwms}.
While this ``lean client'' provides less benefits than more complex ``pilots'', it also belongs to the class of late-binding workload management systems, although at a simpler level.

\subsubsection{Monitoring}
The ability of the user or operator of a WMS to gain immediate and efficient access to information describing the status of jobs, tasks (i.e. collections of jobs) and operations performed on the data
is an essential feature of a good Workload Management System~\cite{pandamon_chep10}. For operators and administrators, it provides crucial debugging and troubleshooting
capabilities. For the users and production managers, it allows better diagnostics of application-level issues and performance, and helps to better plan the user's workflow
~\cite{pandamon_isgc14}.

Of course, all three domains involved in the present discussion (Workflow, Workload, Data) benefit from monitoring capability.

\subsubsection{The Back-end Database}
The power of a successful WMS comes in part from its ability to effectively manage state transitions of the many objects present
in the system (units of data, jobs, tasks etc). This is made possible by utilizing a database to keep the states of these objects.
Most current implementations rely on a RDBMS for this function (e.g. ATLAS PanDA is using the Oracle RDBMS at the time of writing).
The database serves both as the core intrument in the ``brokerage'' logic (matching of workload to resources) and as the source
of data for many kinds of monitoring.

In seeking shared and reusable solutions, we would like to point out that it is highly desirable to avoid coupling of the WMS application code to a particular
type or flavor of the database system, e.g. Oracle vs MySQL etc. Such dependency may lead to difficulties in deployment due to available
expertise and maintenance policies at the target organization and in some cases even licensing costs (cf. Oracle). Solutions such as an ORM layer
or other methods of database abstraction
should be utilized to make possible utilization of a variety of  products as the back-end DB solution for the Workload Management System,
without the need to rewrite any significant amount of its code. This is all the more important in light of the widening application of noSQL
technologies in industry and research, since the possibilities for future migration to a new type of DB must remain open.

\subsection{Intelligent Networks and Network Intelligence}

Once again, the issue of network performance, monitoring and application of such information to improve
throughput and efficiency of workflow belongs to two domains, Workload Management and Data Management. In itself, the network performance monitoring is not a new subject by any means and effective monitoring tools have been deployed at scale~\cite{perfsonar_chep12}. However, until recently, network performance data was not
widely used as a crucial factor in managing the workload distribution in HEP domain in ``near time''. In this approach, network performance information is aggregated from a few sources, analyzed and used in determine a more optimal placement of jobs relative to the data~\cite{panda_chep13}.

A complementary strategy is application of ``Intelligent Networks'', whereby data conduits of specified bandwidth
are created as needed using appropriate SDN software, and utilized for optimized data delivery to the processing location
according to the WMS logic.

\subsection{Opportunities for improvement}


\subsubsection{Focus Areas}
\label{wms_focus}
Technical characteristics of Workload Management Systems currently used in HEP and related fields are regarded
as sufficient (including scalability) to cover a wider range of applications and existing examples support that
(cf. LSST and AMS utilizing PanDA~\cite{pandamon_isgc14}). Therefore, the focus needs to be not on entirely new solutions,
but characterization of the existing systems in terms of reusability, ease of deployment and maintenance, and efficient interfaces.
We further itemize this as follows:

\begin{itemize}

\item \textbf{Modularity} - addressing the issue of the ``Three Domains''~(\ref{3domains}):
\begin{itemize}
\item \textbf{WMS \& Data}: interface of a Workload Management System to the Data Management System needs to be designed in a way that
excludes tight coupling of one to another. This will allow deployment of an optimally scaled and efficient WMS in environments where a pre-existing
data management  system is in place, or where installation of a particular data management system puts too much strain on local resources. For example,
replicating an instance of a high-performance and scalable system like Rucio which is currently used in ATLAS would be prohibitively expensive for
a smaller research organization.

\item \textbf{Workflow Management}: the concept of scientific workflow management is an old one, but recently it's coming to the fore due to
increased complexity of data transformations in many fields of research, both in HEP and in other disciplines. We recommend investigation of both existing
and new solutions in this area, and design of proper interfaces between Workflow Management systems and underlying Workload Management systems.

\end{itemize}

\item \textbf{Pilots}: the technique of deploying Pilot Jobs to worker nodes on the Grid adds robustness, flexibility and adaptability to the system.
It proved very successful at extreme scale and the use of this technique should be encouraged. Creating application-agnostic templates of pilot code
which can be reused by different experiment at scales smaller than LHC could be a cost-effective way to leverage this technique.

\item \textbf{Monitoring}:
\begin{itemize}
\item \textbf{Value}: A comprehensive and easy-to-use monitoring system has a large impact on the overall productivity of personnel operating a Workload Management System.
This area deserves proper investment of effort.

\item \textbf{Flexibility}: The requirements of experiments and other projects will vary, hence the need for flexible and configurable solutions
for the Monitoring System utilized in Workload Management and Data Management.

\end{itemize}

\item \textbf{Back-end Database}: Ideally, there should be no coupling between the WMS application code and features of the specific database system
utilized as its back-end -- this will hamper reusability. Such dependecies could be factored out and abstracted using techniques such as ORM etc.

\item \textbf{Networks}: There are significant efficiencies to be obtained by utilizing network performance data in the workload management process. Likewise,
intelligent and configurable networks can provide optimal bandwidth for workflow execution.

\item \textbf{Cloud}: Workload Management Systems deployed in this decade and beyond must be capable of efficiently operating in both Grid and Cloud environment,
and in hybrid environments as well.

\end{itemize}

\subsubsection{Summary of Recommendations}
\subsubsection*{WMS Inventory}
We recommend that future activities of an organization such as FCE would include an assessment of major existing Workload Management Systems
using criteria outlined in~\ref{wms_focus}, such as
\begin{itemize}
\item modularity, which would ideally allow to avoid deployment of a monolithic solution and would instead allow to
utilize Proper platform and technologies as needed, in the Data Management, Workload Management and Workflow Management domains.
\item Flexibility and functionality of monitoring.
\item Reduced or eliminated dependency on the type of the database system utilized as back-end.
\item Transparency and ease of utilization of the Cloud resources.
\end{itemize}
This assessment will be useful in a few complementary ways:
\begin{itemize}
\item It will serve as a ``roadmap'' which will help organizations looking to adopt a system for processing their workflows in the Grid and Cloud environment,
to make technology choices and avoid duplication of effort.
\item It will help identify areas where additional effort is needed to improve the existing systems in terms of reusability and ease of maintenance (e.g.
implementing more modular design).
\item It will summarize best practices and experience to drive the development of the next generation of distributed computing systems.
\item It may help facilitate the developement of interopability layer in the existing WMS, which would allow future deployment to mix and match components
from existing solutions in an optimal manner.
\end{itemize}
Further, this assessment should also contain a survey of existing ``external'' (i.e. open source and community-based) components that can be utilized
in existing and future systems,  with proper interface design. The goal of this part of the exercise it to identify cases where software reuse may help
to reduce development and maintenance costs. For example, there are existing systems for flexible workflow management which have not been
extensively evaluated for use in HEP and related fields.

It must be reconginzed that due to complexity of the systems being considered, the development of this assessment document will not be a trivial
task and will require appropriate allocation of effort. However, due to sheer scale of deployment of modern WMS, and considerable cost of resources
required for their operation, in terms of both hardware and human capital, such undertaking will be well justified.

\subsubsection*{Cloud Computing}
HEP experiments are entering the era of Cloud Computing. We recommend continuation of effort aimed at investigating and putting in practice
methods and tools to run scientific workflows on the grid. Careful cost/benefit analysis must be performed for relevant use cases.

In addition to extending existing WMS to the Cloud, we must work in the opposite direction, i.e. to maintain efforts to evaluate components of frameworks such as OpenStack~\cite{openstack} for possible ``internal'' use in HEP systems.

\pagebreak
\section{Geometry Information Management}


\subsection{Description}

Almost every HEP experiment requires some system for geometry
information management.  Such systems are responsible for providing a
description of the experiment's detectors (and sometimes their
particle beamlines), assigning versions to distinct descriptions,
tracking the use of these versions through processing and converting
between different representations of the description.  

The geometry description is consumed mainly for the following
purposes:

\begin{itemize}
\item Simulate the passage of particles through the detector or
  beamline.  
\item Reconstruct the kinematic parameters and particle identity
  likely responsible for a given detector response (real or
  simulated).
\item Visualize the volumes to validate the description and in the
  context of viewing representations of detector response and
  simulation truth or reconstructed quantities.
\end{itemize}

The prevailing model for geometry information in HEP is Constructive
Solid Geometry (CSG).  This model describes the arrangement of matter
by the placement of volumes into other volumes up to a top level
``world'' volume.  It is typical to describe this as a daughter volume
being placed into a mother volume.  The placement is performed by
providing a transformation (translation plus rotation) between
conventional coordinate systems centered on each volume.  A volume may have
an associated solid shape of some given dimensions and consist of some
material with bulk and surface properties.  With no such association
the volume is considered an ``assembly'', merely aggregating other
volumes.

While this model is predominant, there is no accepted standard for
representing a description in this model.  Instead, there are a
variety of applications, libraries and tools, each of which makes use
of their own in-memory, transient object or persistent file formats.  

For example, Geant4 is a dominant particle-tracking Monte Carlo
simulation likely used by a majority of HEP experiments.  It defines a
set of C++ classes for the description of a geometry and it can import
a representation of a description in the form of an XML file following
the GDML schema.  It has various built-in visualization methods and
can export to some number of other formats for consumption by others.

Another common example are the \texttt{TGeo} classes provided by ROOT.
These can be constructed directly or via the import of a GDML file
(with some technical limitations).  Like Geant4 objects, ROOT provides
means to track rays through the geometry as well as a few
visualization techniques.

There are stand-alone visualization tools such as HepRApp (which take
HEPREP files that Geant4 can produce), GraXML (which can read GDML
with some limitations or AGDD XML).  There are also CAD programs that
can read OpenInventor files which can be produced.  In experiments
that make use of Gaudi and DetDesc, the PANORAMIX OpenInventor based
visualization library can be used.

\subsection{Unified System}

This variety has lead to a ``tower of babel'' situation.  In practice,
experiments limit themselves to some subset of tools.  Developing
their own solutions is often seen as the least effort compared to
adopting others.  This of course leads to an ever larger tower.  Some
common ground can be had by converting between different
representations.  This is the approach taken by the Virtual Geometry
Model (VGM)\cite{vgm} and the General Geometry Description (GGD)\cite{gdd}.

VGM provides a suite of libraries that allow for applications to be
developed which populate one system of geometry objects (eg, Geant4 or
ROOT).  These can then be converted to a general representation and
finally converted to some end-form.  Care must be taken to keep
implicit units correct and in an explicit system of units (the one
followed by GEANT3).  There is no facilities provided for the actual
authoring of the geometry description and that is left to the
application developer.

Addressing this need for an authoring system is a main goal of GGD.
This system provides a layered design.  At the top is a simple
text-based configuration system which is what is expected to be
exposed to the end user.  This drives a layer of Python builder
modules which interpret the configuration into a set of general
in-memory, transient objects.  These objects all follow a strict CSG
schema which is conceptually compatible with that of Geant4.  This
schema includes specifying an object's allowed quantities and provides
a system of units not tied to any particular ``client'' system.  A
final layer exports these objects into other representations for
consumption by other applications, typically by writing a file.
Access to both the set of general geometry objects and any
export-specific representation is available for access in order to
implement validation checks.  Thus, in GGD the source of geometry
information for any ``world'' is the Python builder modules and the
end-user configuration files.

\subsection{Problems with CAD}

It is not unusual for an experiment at some point to consider
integrating CAD to their geometry information management system.  CAD
provides far better authoring and visualization methods than most HEP
geometry software.  It is typical that a CAD model for an experiment's
detectors or beamline must be produced for engineering purposes and it
is natural to want to leverage that information.  However, there are
several major drawbacks to this approach.

CAD models sufficient for engineering purposes typically contain
excessive levels of information for HEP offline software purposes.
Applied to a tracking simulation this leads to an explosion in the
number of objects that must be queried on each step of the particle's
trajectory.  It leads to large memory and CPU requirements with
minimal or incremental improvements in the quality of the simulation
or reconstruction results.

Use of CAD usually requires expensive, proprietary software with
significant expertise required to use.  This limits which individuals can
effectively work on the geometry and tends to lock in the success of
the experiment to one vendor's offering.  It is typical for the geometry
description to require modification over a significant portion of the
experiment's lifetime and these limitations are not acceptable.

Finally, the use of a CSG model is uncommon in CAD.  Instead a
surface-oriented description is used.  For its use in Geant4, a CSG
model is required.  Converting from surfaces to CSG is very
challenging particularly if the CAD user has not attempted to follow
the CSG model in effect, if not in deed.

There is, however, potential in using CAD with HEP geometries.  This
is being explored in GGD in the production of OpenInventor files which
can be loaded into FreeCAD, a Free Software CAD application.  While
FreeCAD can currently view an OpenInventor CSG representation it can
not be used to produce them.  However, FreeCAD is extensible to new
representation types.  With effort, it may be extended to produce and
operate on a suite of novel CSG objects which follow a schema
similarly to that required by Geant4.

\subsection{Opportunity for improvement}

The ``tower of babel'' situation should be addressed by putting effort
in to following areas:

\begin{itemize}
\item Form a small working group from geometry system experts to
  develop a formal data schema to describe the CSG objects that make
  up a general geometry system.  This schema should be independent
  from any specific implementation but be consistent with major
  existing applications (specifically Geant4 and ROOT).  The schema
  should be presented in a generic format but made available in a form
  that can be directly consumed (eg, JSON or XML) by software.
\item A general, transient data model for use in major programming
  languages (at least C++ and Python) should be developed which
  follows this schema.  Independent and modular libraries that can
  convert between this data model and existing ones (GDML, ROOT)
  should be developed.  One possibility is to further develop VGM in
  this direction.
\item Develop a general purpose geometry authoring system that can
  produce objects in this transient data model.
\end{itemize}

\pagebreak
\section{Conditions Databases}


\subsection{Description}

Every HEP experiment has some form of ``conditions database''. The purpose of such as database is to capture and store any information that is needed in order to interpret or simulate events taken by the DAQ system. The underlying principle behind such a database is that the ``conditions'' at the time an event is acquired vary significantly slower that that the quantities read out by the DAQ in the event itself. The period over which these condition quantities can change range from seconds to the lifetime of the experiment.

In implementing a conditions database, an experiment is providing a mechanism by which to associate an event to a set of conditions without having to save a complete copy of those conditions with every event. A secondary feature is that the event-to-conditions association can normally be configured to select a particular version of the conditions as knowledge about the conditions can change over time as they are better understood.

\subsection{Basic Concepts}

It turns out that the basic concepts of a conditions database do not vary between experiments. They all have the same issues to solve. Questions of scale, distribution and so forth can depend on the size and complexity of the data model used for quantities within the database, but these aspect are secondary and are addressed by the implementation. The resulting software for experiment differ more in the choices of technologies used rather than any conceptual foundation.

\subsubsection{Data Model}

The Data Model of a conditions database defines how information is grouped in atomic elements in that data base and how those atomic elements are structure so that clients can recover the necessary quantity. This is the most experiment specific concept as it is directly related to the object model used in the analysis and simulation codes. However the division of condition quantities into the atomic elements is normally based on two criteria.

\begin{itemize}
\item The period over which a quantity varies, for example geometry may be updated once a year, while a detectors calibration may be measured once a week.

\item The logical cohesiveness of the quantities, for example the calibrations for one detector will be separate from those of another detector even if they are updated at the same frequency.
\end{itemize}

\subsubsection{Interval of Validity}

The standard way of matching a conditions element to an event is by using a timestamp related to the event's acquisition. Given this time the conditions database is searched for the instance of the element that was valid at that time. (What to do when multiple instances are valid for a given time is dealt with by versioning, see section~\ref{conditions-versioning}.). This therefore requires each entry in the conditions database to have an interval of validity stating the beginning and end times, with respect to the events timestamp, for which it should be considered as the value for its quantity.

As analysis often proceed sequential with respect to events, most implementations of condition database improve their efficiency by caching the `current' instance of a quantity once it has been read from the database until a request is made for a time outside its interval of validity. At this point the instance appropriate to the new time will be read in, along with its interval of validity.

\subsubsection{Versioning}
\label{conditions-versioning}

During the lifetime of an experiment a database will accumulate more than one instance of a conditions element that are valid for a given time. There two most obvious causes of this are the following.

\begin{itemize}
\item A conditions element is valid from a given time to the end-of-time in order to make sure there is always a valid instance of that element. At a later time during the experiment a new value for the element is measured and this is now entered into the database with its interval of validity starting later than the original instance but, as it in now the most appropriate from there on out, its validity runs until the end-of-time as well.

\item A conditions element may consist of a value derived from various measurements. In principle this can be considered a `cached' result of the derivation however it is treated as a first class element in the conditions database. At some point later, a better way of deriving the value is developed and this new value is placed in the database with the same interval of validity as the orignal one.
\end{itemize}

In both cases there need to be a mechanism for arbitrating which instances are used. This arbitration is managed by assigning versions to each instance. The choice of which version to used depends on the purpose of job that is being executed. If the purpose of the job is to use the `best' values then the `latest' version is used, but if the purpose of the job is to recreate the results of a previous job or to provide a know execution environment then it must be possible to define a specific version to be used by a jobs.

In order for the above versioning to work there must be some monotonic ordering of the versions. Typically this is done by the `insertion' date which is the logical date when the version was added to the database. It should be noted here that this date does not always reflect the actual date the version was inserted as that may not create the correct ordering of versions.

\subsection{Examples}

The following is the subset of the implementations of a conditions database pattern already done by HEP experiments.

\subsubsection{DBI, Minos}

This is a C++ binding that is decoupled for its surrounding framework, a feature that allowed it be adopted by the Dayabay Experiment for use in its Gaudi customization. It has a very thin data model with a conditions item being a single row in an appropriate SQL table.

\subsubsection{IOVSvc, Atlas}

The Atlas Interval of Validity Service, IOVSvc, is tightly bound to its Athena framework (their customization of the Gaudi framework). It acts by registering a proxy that will be filled by the service rather than direct calls to the service. It all has a feature where a callback can be registered that is invoked by the service whenever the currently value conditions item is no longer valid.

\subsubsection{CDB, BaBar}

The BaBar conditions database is notable in that during its lifetime it went through a migration for an ObjectivityDB back end to on using RootDB. This means that it can be used as an example of migrating conditions database implementations as suggested in the next section.

\subsection{Opportunity for improvement}

Given that the challenge of a condition database, to match data to an events timestamp, is universal and not specified to any style of experiment, and given that numerous solutions to this challenge exist, there is little point in creating new ones. The obvious opportunities for improvement are ones that make the existing solutions available for use by other experiment in order to avoid replication (again). To this end the following approaches are recommended:

\begin{itemize}
\item A detail survey of the interface of existing solutions should be made. The result of this survey should the definition of a ``standard'' API for conditions databases that is decoupled for any particular framework. This standard would be a superset of the features found in existing implementations so that all features of a given implementation can be accessed via this API. Suitable language bindings, such as C++ and python, so be provided as part of the standard.

\item Given the standard API and language bindings provided by the previous item, maintainers of conditions database implementations should first be encouraged, where possible, to develop the necessary code to provide the API as part of their implementation. They should then be encouraged to extent their own implementation to cover as much of the API as it is possible for their technology to support.

\item On the `consumer' side of the API, existing framework maintainers should be encouraged to adapt their framework so that is can use the standard API to resolved conditions database access. For frameworks and conditions databases that are tightly couples, such as the Gaudi framework and its IOVsrc, this item, in concert with the previous one, will enable the conditions database to be decoupled for the analysis code.

\item In the longer term, given the standard conditions database API, the development of a Software-as-a-Service for conditions databases, for example use a RESTful interface, should be encouraged. This would allow the provisioning of conditions databases to be moved completely out of the physicist's realm and into that of computing support with is more suited to maintain that type of software.
\end{itemize}

\pagebreak


\end{document}